\documentclass[aps,prl,preprintnumbers,twocolumn,groupedaddress,footinbib]{revtex4-1}

\pdfoutput=1

\usepackage{amsmath,amssymb,mathrsfs}
\usepackage{slashed}
\usepackage{xcolor}
\usepackage[colorlinks=true,linkcolor=blue,citecolor=blue,urlcolor=violet]{hyperref}
\usepackage{soul}
\usepackage{tikzfeynman}
\usepackage{ulem}

\newcommand{\be}{\begin{equation}}
\newcommand{\ee}{\end{equation}}
\newcommand{\Mpc}{\mathrm{Mpc}}

\newcommand{\Mpl}{M_{\rm Pl}}

\begin{document}

\title{Stochastic Dark Matter from Curvature Perturbations} 

\author{Raghuveer Garani}
\email{garani@iitm.ac.in}
\affiliation{Department of Physics, Indian Institute of Technology Madras, Chennai 600036, India}

\author{Michele Redi}
\email{michele.redi@fi.infn.it}
\affiliation{
INFN Sezione di Firenze, Via G. Sansone 1, I-50019 Sesto Fiorentino, Italy
}
\affiliation{Department of Physics and Astronomy, University of Florence, Italy
} 

\author{Andrea Tesi}
\email{andrea.tesi@fi.infn.it}
\affiliation{
INFN Sezione di Firenze, Via G. Sansone 1, I-50019 Sesto Fiorentino, Italy
}
\affiliation{Department of Physics and Astronomy, University of Florence, Italy
}

\begin{abstract}
We investigate the production of dark matter from curvature perturbations produced during inflation or
in standard cosmology, for example during first order phase transitions. Perturbations break Weyl flatness 
of the Friedmann-Lemaitre-Robertson-Walker metric, allowing conformally coupled fields -- in particular fermions studied here --  to be produced even in the massless limit. Particle production can be computed by studying the Bogoliubov transformation induced by the stochastic background.
For perturbations generated during inflation, we present a closed formula for the resulting abundance of particles that depends 
solely on the power spectrum of curvature perturbations at the end of inflation. This production mechanism 
can be dominant especially if the amplitude of curvature perturbations is enhanced for modes that exit the horizon
towards the end of inflation. In the simplest scenario, the critical dark matter abundance is reproduced for $M \gtrsim 10^{6}$ GeV. 
\end{abstract}
\maketitle

\paragraph{\bf Introduction.--}
Stochastic perturbations generated by inflation are widely believed to be the seeds of structures that 
we observe in the universe today, as confirmed by the measurements of the Cosmic Microwave Background (CMB) and large scale structure of the universe.

It is intriguing to speculate that the very same perturbations might also be responsible for the origin of Dark Matter (DM), exploiting ingredients that are common both to gravitational particle production in the expanding universe and to the formation of primordial black holes from inflation.
In this Letter we will show that any particle DM candidate - described by a quantum field -  has an unavoidable gravitational production mode just based on the existence (say the presence) of perturbations, focusing on scalar perturbations originating from inflation.  The same mechanism was first noticed for gravitational wave (tensor) perturbations from phase transitions and other sources in Refs.~\cite{kopp1,kopp2}. Our result will provide a one-to-one correspondence between the DM abundance today and the stochastic curvature power spectrum that sets the initial conditions of our Universe. 

We emphasize that gravitational production of DM -- or any other dark sector -- is an unavoidable consequence of Weyl invariance of conformally coupled fields.  The plot is as follows: The Friedmann-Lemaitre-Robertson-Walker (FLRW)  metric describes an isotropic and homogenous - albeit expanding - universe. Since the metric is flat up to a Weyl rescaling, particle production only occurs when the mass becomes relevant as this is the only source of breaking Weyl invariance at the classical level. This happens when the mass becomes comparable with the Hubble parameter and can be studied using Bogoliubov transformations between initial and final vacua; see the seminal works~\cite{Ford:1986sy,Chung:1998zb} and~\cite{Kolb:2023ydq} for a review. In contrast, massless particles - except for Nambu-Goldstone bosons and gravitons - are never produced in a FLRW metric, as Weyl invariance effectively makes them propagate as in flat space.

The situation changes drastically when perturbations are added to the FLRW background, as they cannot be reabsorbed by a Weyl transformation, allowing for the production of any particle even in the massless limit. Notably, this also permits the production of fermions and gauge fields. In this case, particles can be produced with an abundance that is quadratic in the perturbations, i.e., linear in the power spectrum of inflationary fluctuations. 

\medskip
In the simplest version studied in this Letter, the  scenario is realized by a Weyl fermion singlet $\psi$ - with mass $M$ - that lives in our universe, secluded from the SM. Using 2-component Weyl fermions \cite{Wess:1992cp}, the theory is described by
\begin{equation}\label{model}
\mathscr{L} = i \bar{\psi} {\bar \sigma}^\mu (\partial_\mu+\omega_\mu) \psi - \frac M 2 (\psi\psi +\bar{\psi} \bar{\psi})\,,
\end{equation}
where $\bar{\sigma}^\mu\equiv (1, -\vec{\sigma})$ and $\omega_\mu$ is the spin connection necessary to couple the fermion to a curved background.

If we neglect possible direct decay of the inflaton to $\psi$ and other model-dependent contributions, 
two unavoidable production mechanisms of DM exist related to gravity $i)$ gravitational freeze-in (GFI), which extracts energy from the SM thermal bath~\cite{Garny:2015sjg}, and $ii)$ gravitational particle production (GPP) in FLRW background (see~\cite{Kolb:2023ydq} for a review). In the first case, the number density rapidly decreases with the reheating temperature as $\sim (T_R/M_{\rm Pl})^3$. In the second case, the number density is suppressed by $M$. As a consequence the abundance of DM is reproduced for  large masses, $M> 10^{8}$ GeV. 

The new idea presented in this Letter is the possibility of producing unsuppressed number densities of particles exploiting the presence of primordial stochastic gravitational fluctuations on top of the FLRW background metric. Therefore we dub this new mechanism ``Stochastic  particle production".
On the technical side we will demonstrate how to derive the particle abundance in this case using the formalism of Bogoliubov transformations, generalizing the computation in an expanding background.

\medskip
\paragraph{\bf Fermion production in an FLRW universe.--}
To set the stage and introduce our notation, we briefly review particle production from quantum fluctuations in an expanding universe.
In FLRW spacetime, using conformal time, the metric takes the form $g_{\mu\nu}=a(\tau)^2 \eta_{\mu\nu}dx^\mu dx^\nu$, that is flat up to Weyl rescaling. 
The time dependence implies, in general, that energy is not conserved, leading to a time-dependent Hamiltonian in quantum mechanics. Therefore, if the system is in the ground state at some initial time in the far past, this will be interpreted as an excited state from the point of view of an observer at late time.
The mapping between the Hilbert space at early and late times is known as a Bogoliubov transformation. 

Let us now discuss the Weyl fermion. Defining $\psi\equiv \chi/a(\tau)^{3/2}$, the equation of motion takes the form
\be\label{eq:Dirac}
i\bar\sigma^\mu \partial_\mu \chi =  M a \bar{\chi}\,.
\ee
As a consequence of Weyl invariance, in the limit $M=0$, this equation describes a fermion in flat space, resulting in 
no particle production. In this case the quantized field is
\be
\chi_0(\tau,\vec x)=\int \frac{d^3k}{(2\pi)^3}\bigg[ u_{\vec k}(\tau) e^{+i \vec k \cdot \vec x} a_{\vec k} +  v_{\vec k}(\tau) e^{-i \vec k \cdot \vec x} b^\dag_{\vec k}\bigg],
\ee
where the spinor wave-functions are the usual positive and negative frequency solutions. In particular
\be
\label{eq:flat_u}
u_{\vec k}(\tau)=\xi_{\vec k}^{-}\, e^{- i k\tau}\,,
\ee
and similarly for $v_{\vec k}(\tau)$. We have also introduced $\xi^\pm_{\vec k}$ as eigenstates of the helicity operator $\hat h_{\vec k}\equiv \vec \sigma \cdot \vec k/|k|$. The creation/annihilation operators $a_{\vec k}$ and $b_{\vec k}$ have canonical anti-commutation relations, and they annihilate the vacuum state. As usual  negative/positive frequencies are associated to creation/annihilation operators.

\medskip
The mass instead explicitly breaks Weyl invariance, causing the equation to depend explicitly on time and leading to particle production. The situation is a tad more complicated than for scalars due to the spinor structure, see ~\cite{Ema:2019yrd} and the appendix of~\cite{jump}.  It is now convenient to diagonalize \eqref{eq:Dirac}, introducing $\chi^\pm\equiv\chi \pm \bar\chi$, that solve  $\Box \chi^\pm + (M^2 a^2 \mp i M a')\chi^{\pm}=0$, with the constraint given by the Dirac equation. In flat space at low momenta one finds $\chi^{\pm}\sim e^{\mp i M \tau}$, corresponding to the creation and annihilation of massive fermions. 

In the expanding universe positive, the frequency mode in eq.~\eqref{eq:flat_u} evolves into a combination
of positive and negative frequency waves at late times. Asymptotically, in the future we have
\begin{equation}
\chi^-_{\vec k}(\tau) \sim  \alpha_{\vec k}(\tau) \xi_{1} e^{-i \int^\tau \omega_{k}(\tau') d\tau'} +  \beta_{\vec k}(\tau) \xi_{2} e^{i \int^\tau \omega_{ k}(\tau') d\tau'} \,,
\label{eq:alarming}
\end{equation}
where $\omega_{ k}^2=k^2 +M^2 a^2$ where $\xi_{1,2}$ are constant spinors.
The presence of negative frequencies signals that particles are produced. 
One can show that
the creation/annihilation operators of the initial vacuum  are a linear combination of the ones at late times. They are related by a Bogoliubov transformation with parameters $\alpha_{\vec k}$ and $\beta_{\vec k}$, where $|\alpha_{\vec k}|^2+|\beta_{\vec k}|^2=1$.

The parameter $\beta_{\vec k}$ in eq. \eqref{eq:alarming} is precisely the  coefficient of the Bogoliubov transformation. 
From this one determines the number density of particles produced at late time where the curvature of spacetime is negligible as
\be\label{dndk}
\frac{d (na^3)}{d\log k} = \frac{k^3}{2\pi^2}|\beta_{\vec k}(\tau\to\infty)|^2\,.
\ee

\medskip
\paragraph{\bf Particle production from perturbations.--} 
Let us now  consider the case where scalar fluctuations are added on top of the homogeneous FLRW background.
These perturbations explicitly break the Weyl form of the metric, and thus particle production is expected even in the limit $M=0$. 
This effect can be computed generalising the Bogoliubov transformation above in the presence of scalar metric fluctuation, 
as we now describe. The result is particularly simple because the computation to leading order can be performed in flat spacetime. 
Further details are provided in the companion paper~\cite{future}.

We work in conformal Newtonian gauge, where the line element has the form
\be\label{eq:conformal-gauge}
ds^2 = a^2 d\tau^2[1+2\Phi(\tau,\vec x)] - a^2 d\vec{x}^2 [1-2\Psi(\tau, \vec x)]\,.
\ee
For simplicity we consider the case where no anisotropic stress is present in the perturbations, leading to $\Psi=\Phi$. 
Upon   Weyl rescaling,  the fermion field $\chi$ satisfies the Dirac equation
\be\label{equation-pert}
i\bar\sigma^\mu \partial_\mu \chi = i \left[2  \Psi \dot\chi - \frac 1 2(\nabla \Psi)\cdot (\vec \sigma\chi)+\frac 3 2 \dot\Psi\chi\right]\,.
\ee
It is very tempting now to draw an analogy with eq.~\eqref{eq:Dirac}, since on the right-hand side of eq.~\eqref{equation-pert} we have a time-dependent source linear in the fermion field. However, there are a few technical differences. First, helicity is conserved - since the fermion is now massless. Second, the source breaks both time and spatial translational invariance due to the space-time dependence of $\Psi$; and the third is that the source is stochastic in nature. Helicity conservation is not instrumental in what follows, although it simplifies our formulae. The other two technical differences are important: the breaking of translations will mix modes with different $\vec k$, while the stochastic nature of the source forces us to consider averages.

To determine the Bogoliubov transformation induced by the perturbations, we need to solve eq.~\eqref{equation-pert}
with positive frequency initial conditions at early time as in eq.~\eqref{eq:flat_u}.  Working in Fourier space, 
eq.~\eqref{equation-pert} can be cast in the form
\be\label{MS-source}
 i\partial_\tau \chi_{\vec k} + (\vec{k} \cdot \vec{\sigma}) \chi_{\vec k} = J_{\vec k}(\tau)\,.
\ee
where the source $J_{\vec k}$ is given by
\be
J_{\vec k}(\tau)=  \int \frac{d^3q}{(2\pi)^3} \left[2  \Psi_{\vec q} \dot\chi_{\vec \omega}  - \frac i 2 \Psi_{\vec q}\, (\vec q\cdot  \vec \sigma)\chi_{\vec \omega}+\frac 3 2 \dot\Psi_{\vec q}\chi_{\vec \omega}\, \right]~.
\ee
Here and in the following $\vec \omega = \vec k - \vec q$,  and we define  $x\equiv \cos \theta=\vec k \cdot \vec q/(kq)$.

This equation can be easily solved perturbatively in $\Psi$.  
The zeroth order solution $\chi_{0,\vec k}$ -- obtained by neglecting the source -- is again given by eq.~\eqref{eq:flat_u}.
Plugging this on the right-hand side the source is a known function $J_{\vec k}(\chi_0)$. 
To first order the correction to the wave-function can be conveniently obtained using the retarded Green function,
\be\label{eq:Greens1}
\chi_{\vec k}(\tau)= \chi_{0,\vec k}(\tau)+ \int_{-\infty}^\tau d\tau' G_{\vec k}^R(\tau,\tau') \chi_{0,\vec k}(\tau')\,.
\ee
The explicit expression for $G_{\vec k}^R(\tau,\tau')$ is found to be
\be\label{eq:Greens} 
G_{\vec k}^R(\tau,\tau')=i\theta(\tau-\tau') \left[ P^{-}_{\vec k} e^{-i k (\tau-\tau')} + P^{+}_{\vec k}e^{i k (\tau-\tau')}  \right]\,,
\ee
where $P^\pm_{\vec k}= \frac12 (\mathbf{I}\pm \hat h_{\vec k})$ are the helicity projectors. 

Since the $G_{\vec k}^R(\tau,\tau')$ is expanded in positive and negative frequencies we can immediately extract the Bogoliubov coefficient $\beta_{\vec k}$
from the negative frequency wave.   The Bogoliubov coefficient $\beta_{\vec k}$ is extracted by projecting onto the negative frequency solution $e^{+i k \tau}$ which is associated to the positive helicity spinor. Therefore it is explicitly given at $\tau\to \infty$ by
\be
\beta_{\vec k}(\tau) = \int_{-\infty}^\tau d\tau' e^{-i (k+\omega) \tau'}  \xi_{\vec k}^{+\,\dag}  P^{+}_{\vec k} J_{\vec k}(\tau')\,.
\label{eq:beta}
\ee
The formula above is valid  for any classical background to leading order in the perturbation.
Since $\beta_{\vec k}$ is generated at first order in $\Psi$ it follows that particle number \eqref{dndk} will be quadratic.

In this work we will be interested in stochastic background characterised by a 2-point function at unequal times given by
\begin{equation}\label{eq:power-spectrum}
\langle \Psi_{\vec q}(\tau) \Psi^*_{\vec q'}(\tau') \rangle= (2\pi)^3 \delta^3 (\vec{q}-\vec{q'}) \frac{2\pi^2}{q^3}\Delta_\Psi(q,\tau,\tau'),
\end{equation}
Using eq. (\ref{eq:beta})  and the assumption of stochastic background (\ref{eq:power-spectrum}) we can derive the expectation value of $|\beta_{\vec k}|^2$
that determines the number of particles produced. One finds,
\begin{equation}\label{master-formula}
\begin{split}
\langle |\beta_{\vec k}|^2 \rangle&= \frac 1 2  \int d\tau \int d\tau' \int  d(\log q)\int dx  \,\\
& \times e^{-i (k+\omega)(\tau-\tau')}\Delta_\Psi(q,\tau,\tau') {\cal K}[k\,, q\,, x ] \,.
\end{split}
\end{equation}
We have here introduced the kernel $\mathcal{K}$,  
\begin{equation}
 {\cal K}[k\,, q\,, x]= -\frac{(k-\omega )^2 \left(k^2-2 k \omega -q^2+\omega ^2\right)}{4 k \omega }\,.
\end{equation}
This formula is analogous to the one for production from gravity waves derived in \cite{kopp1,kopp2} using the in-in formalism. 
Note that differently from ordinary particle production for a stochastic background $\langle\beta_{\vec k}\rangle=0$ but $\langle |\beta_{\vec k}|^2\rangle\ne 0$.

In order to derive this formula we have integrated by parts the contribution proportional to $\dot{\Psi}$ to relate  it to the one of $\Psi$.
The boundary term can be neglected if the integration extends to sufficiently early times where the source vanishes.
The explicit time dependence is encoded in the power spectrum of $\Psi(x\,,\tau)$. One can check for example 
that no particle production takes if the background is static, as expected from energy conservation. 
 
This formula is one of the main novelties of our work, as it allows us to determine the abundance and distribution 
of particles from the power spectrum in terms of a kinematical kernel. 
A completely analogous formula can be derived for fields of different spins produced from scalar or gravitational wave backgrounds~\cite{future}. 

\medskip
\paragraph{\bf Stochastic DM from curvature perturbations.--} 
With eq.~\eqref{master-formula}, we can proceed to explore several phenomenological scenarios. 
Many physical mechanisms generate a power spectrum as in eq.~\eqref{eq:power-spectrum}, for example first order phase transitions.
However, within the SM we already have a source that we can exploit: the scalar power spectrum of inflationary fluctuations that provides the initial conditions of our universe.

In the remainder of this Letter, we study DM produced from adiabatic scalar fluctuations produced during inflation. 
To define our cosmological scenario, we assume standard slow-roll inflation for modes tested by the CMB.
At smaller scales, corresponding to modes that exit the horizon later during inflation the scalar power spectrum is very poorly known
and could be large while still being compatible with current constraints. This can lead to interesting phenomena, such as the formation
of primordial black holes~\cite{Ozsoy:2023ryl} and secondary gravity waves~\cite{Domenech:2021ztg}. 
Specifically, we consider a power spectrum for the quantum scalar fluctuations $\zeta(\tau, \vec x)$ (see~\cite{Ma:1995ey} for conventions), with  $\Delta_\zeta(q)$ that can be sizeable - say $\Delta_\zeta\approx 10^{-3}$ - for a selected ranges of modes with $q_*\gg q_0$ where $q_0$ is the CMB pivot scale of $0.05/\Mpc$. In this notation, the power spectrum is $\Delta_\zeta(q_0)=2.1 \times 10^{-9}$ with a small tilt~\cite{PLANCK}.  We assume single-field inflation, which leads to a constant $\zeta$ on super-horizon scales. In this situation, $\Delta_\zeta(q)$ is computed once for all at the end of inflation, which we take to be at $\tau=0$.

We also allow for a finite duration of reheating in our cosmological scenario, from the end of inflation until conformal time $\tau_R$, where the universe is reheated at $T_R$.  As customary in standard cosmology, the super-horizon value of $\Psi$ is matched to that of $\zeta$. The matching is done at some early time $\tau$ when the mode is super-horizon for all relevant $k$. This procedure is also adopted here to compute the power appearing in eq.~\eqref{master-formula}. We have
\be\label{inflationary-PSI}
\Delta_\Psi(q,\tau,\tau')= T(q,\tau)\,T(q,\tau') \Delta_\zeta(q)\,
\ee
where $T$ is the transfer function for the mode $\vec q$ from the end of inflation to any later conformal time $\tau$. Notably, for modes that re-enter the horizon in radiation or matter dominance, the transfer function evaluates to $-2/3$ and $-3/5$ respectively as $\tau\to0$. 
The transfer functions here are the ones of standard cosmology, for example in radiation domination $T(k,\tau)=2 \left[\cos(c_s k\tau)-\sin(c_sk\tau)/(c_s k\tau)\right]/(c_s k\tau)^2$, where $c_s\approx 1/\sqrt{3}$ is the sound-speed of the baryon-photon fluid.  In general cosmologies - if needed - they can be computed numerically with codes such as \texttt{CLASS}~\cite{CLASS}. Moreover the $\Psi=0$ during inflation so that $T(q,\tau)=0$ for $\tau<0$.

We now have all the ingredients, and we can compute the DM abundance. By including eq.~\eqref{inflationary-PSI} in our master formula~\eqref{master-formula}, all integrals can be performed. 

Before providing a numerical study, let us simplify the formula. We notice that the time integrals only affect the transfer functions and it is convenient to define 
\be
\mathcal{I}(q,k+\omega)\equiv \int_{-\infty}^\infty d\tau \,e^{-i(k+\omega)\tau}T(q,\tau)\,.
\ee
This function -- the time Fourier transform of the transfer function -- depends in general upon the cosmological scenario under consideration.

The differential numerical abundance of particles produced  by curvature perturbations generated by inflation is thus
\begin{equation}
\frac{d(na^3)}{d\log k}=  \frac{k^3}{4\pi^2} \int d(\log q)\,dx\, \Delta_\zeta(q)   |\mathcal{I}(q,k+\omega)|^2 {\cal K}[k\,, q\,, x] \,.
\label{eq:energyspectrum}
\end{equation}
that depends on the particle type through the kernel $\mathcal{K}$, on the cosmological evolution through $\mathcal{I}$ and on the primordial power spectrum $\Delta_\zeta$. 

In the case of instantaneous reheating or more generally for modes that re-enter the horizon during radiation domination the formula above can be further simplified.
The DM abundance today can be computed performing the integral in $k$ and $\cos\theta$, to get
\be\label{formula2}
n a^3= \frac{A}{4\pi^2} \int d(\log q)\, q^3 \Delta_\zeta(q)\,,~~~~~~~A\approx 0.015
\ee
We emphasize that the integral $A$ that controls the abundance is finite due to the fact that ${\cal K}\propto q^4/k^2$ at large $k$.
Remarkably, the numerical abundance is fully determined by the power spectrum of curvature perturbations. 
This result could have been anticipated: in the limit of vanishing mass, the production is linear in the power spectrum and the only dimensionful scale is given by $q$, so the 
result must depend only on $q^3 \Delta_\zeta(q)$.  

As we will discuss below, the scenario with slow reheating, which we approximate as a phase of matter domination, is particularly interesting.
In this case, the only modification to eq.~\eqref{eq:energyspectrum} is through the transfer function $\mathcal{I}(q,k+\omega)$, which also depends on the reheating time $\tau_R$.
Since in this phase the gravitational potentials are constant, there is no particle production until reheating. While eq.~\eqref{formula2} must be slightly modified, we find however that the total abundance is almost unchanged, see \cite{future} for more details. 

\begin{figure}
\begin{center}
\includegraphics[width=\linewidth,height=\linewidth]{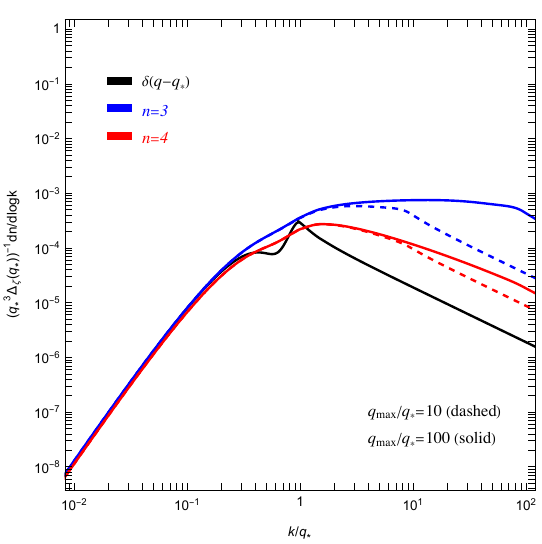}~\\
\caption{ \label{fig:spectra}\small Differential number density of particles eq.~\eqref{eq:energyspectrum} produced by perturbations that re-entered the horizon in radiation domination. Different inputs for $\Delta_\zeta(q)$ are considered, as broken power laws (and Dirac delta) with a peak at the scale $q_*$. Different values for the ratio $q_{\rm max}/q_*$ are also shown.}
\end{center}
\end{figure}

In figure~\ref{fig:spectra}, we show the energy spectrum for different choices of the primordial power spectrum:
a $\delta$-function at the scale $q_*$ (black curve) and a broken power law peaked at $q_*$ of the form $\Delta_\zeta\propto 2(q/q_*)^3/(1+ (q/q_*)^{3+n})$. Let us mention that the UV part of the primordial spectrum $\Delta_\zeta(q)$ has a physical cut-off at $q_{\rm max}$, corresponding to the wave-number of the mode that exited the horizon at around the 60th e-fold of inflation. Therefore, we compute the DM number density integrating up to $q_{\rm max}$, for different choices of $q_{\rm max}/q_*=[10,100]$. 
From eq. \eqref{formula2} for $n>3$ the spectrum is dominated at $q_*$ while in the opposite regime it is peaked at $q_{\rm max}$.

\medskip
\paragraph{\bf Stochastic Dark Matter  Phenomenology.--}~\\ 
From the numerical abundance in eq.~\eqref{formula2}, it is then straightforward to compute the DM abundance $\Omega_{\rm DM}$ today. Being the number density conserved (modulo the expansion of the universe), when the particles produced by stochastic perturbations become non-relativistic, they will contribute to the DM abundance. For a power spectrum peaked at comoving scale $q_*$, the abundance is obtained from eq.~\eqref{formula2},
\be
\Omega_{\rm DM}|_{\rm stochastic}\approx \frac{A}{4\pi^2} \frac{ M\,   q_*^3}{3 \Mpl^2 H_0^2}\Delta_\zeta(q_*)\,.
\label{eq:Omega_sto}
\ee
This formula is strictly derived assuming radiation domination but as we mentioned the result is very similar for modes that re-enter the horizon during reheating. 

Demanding that eq.~\eqref{eq:Omega_sto} reproduces the DM abundance $\Omega_{\rm DM}\approx 0.25$, we can determine the peak of the power spectrum as
\be
q_*\approx  10^{-7} \mathrm{eV} \left(\frac {10^6\,{\rm GeV}}M \right)^{\frac13} \left(\frac {0.01}{\Delta_\zeta(q_*)} \right)^{\frac13}  \left(\frac {0.01}{A}\right)^{\frac13} \,.
\ee
Notice that such comoving momenta correspond to modes that exit the horizon at around a number of e-folds $N =  \log(q_*/H_0)\approx 60$,
suggesting that the larger amplitude may be connected to the end of inflation.

\begin{figure}
\begin{center}
\includegraphics[width=1.15\linewidth, height=1.15\linewidth]{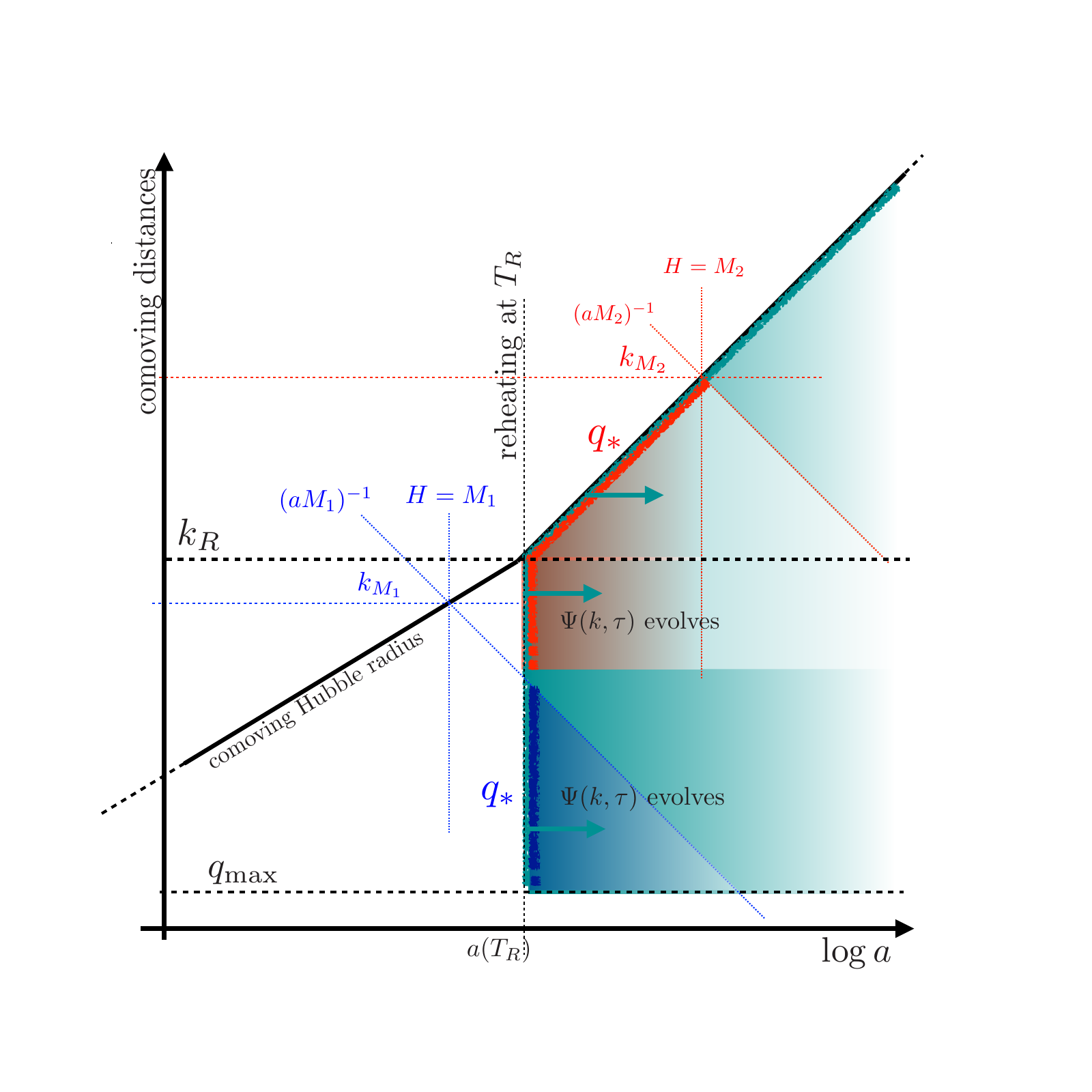}~\\
\caption{ \label{fig:cartoon}\small Contributions to the production for different mass values $M_{1,2}$ corresponding to $k_M>k_R$ or $k_M<k_R$. We compare the comoving Hubble radius with the relevant quantities allowing for a finite duration of reheating (matter domination). We assume production to be dominated by modes $k\approx q_*$ as shown in figure \ref{fig:spectra}. Our stochastic production proceeds when the field can be considered massless, that is for all momenta $k\sim q_* > a M_{1,2}$ respectively, otherwise GPP dominates. Stochastic production occurs when the gravitational potential $\Psi$ starts to evolve in time after reheating (green region) and it is maximal at the blue and red edges respectively for $M_1$ and $M_2$.}
\end{center}
\end{figure}

\medskip
Several comments are in order that we summarize  in Fig. \ref{fig:cartoon}.
First, not surprisingly from eq.~\eqref{eq:energyspectrum} we notice that if the curvature power spectrum is peaked at $q_*$, the DM abundance produced will be peaked at $k\sim q_*$, at least if the power spectrum is sufficiently peaked. This is also shown in Fig.~\ref{fig:spectra}. Therefore it makes sense to use $q_*$ as our reference wave-number. 

Second, since the energy available in perturbations is $q_*/a$, the stochastic production of particles is only efficient if the mass is negligible.  
The massless computation is thus valid as long as $k\sim q_*\gtrsim a M$ at the production time. Due to energy conservation, particle production only takes place when $\Psi$ starts to evolve. We recall that $\Psi$ is constant in time on super-horizon scales $q< a H$, and it remains constant on sub-horizon scales during reheating. Therefore, the first significant time evolution of $\Psi$ happens after $T_R$  (green region in the Fig.~\ref{fig:cartoon}), and we denote $k_R \approx T_0 T_R/\Mpl$ the mode that re-enters at this epoch. For $q_*\gtrsim k_R$, the evolution of $\Psi$ initiates at $T_R$ while in the opposite regime it starts when $q_*\approx a H$. 

Last, another important scale is the comoving momentum $k_M$ that re-enters the horizon when Hubble crosses the mass, i.e.  $k_M/a=H=M$. If this happens in radiation, $q_*>k_M$ is sufficient to guarantee that the mass is negligible (red region in the Fig.~\ref{fig:cartoon}). On the contrary if $k_M$ re-enters the horizon during reheating, $q_*$ needs to be larger than $k\sim q_*\gtrsim a(T_R) M\approx T_0 M/T_R$ for the fermion to be  effectively massless while $\Psi$ evolves (blue region in the Fig.~\ref{fig:cartoon}).

\medskip
\paragraph{Comparison to other mechanisms.--}~\\
Let us now compare our mechanism with GPP induced by the mass $M$ in the expanding universe. 
The abundance from GPP can be computed precisely as described in~\cite{Ema:2019yrd,jump},
\be
\Omega_{\rm DM}|_{\rm GPP} \approx 10^{-2} \frac{ M\,   k_M^3}{3 \Mpl^2 H_0^2}\times \eta\,,
\label{eq:Omega_GPP}
\ee
where $\eta=\mathrm{min}\left[1\,,\frac {T_R}{\sqrt{M\,\Mpl}}\right]$ is a dilution factor different from one if the production happens during reheating.
This formula is strikingly similar to eq.~\eqref{eq:Omega_sto} with an important difference. 
In GPP the peak of the spectrum arises at $k_M$ that corresponds to the largest deviation from adiabaticity of the wave equation due to the mass.
This is analogous to the stochastic particle production where however the non adiabaticity is not order 1 but proportional to $\Delta(q_*)$.
A remarkable difference is instead the dilution factor $\eta$ that suppresses the abundance of GPP during reheating, corresponding to  $k_M>k_R$, due to the entropy injection in the SM thermal bath. On the contrary, our production mechanism is not suppressed by reheating, as the production takes place at $T_R$ or later.
This allows stochastic particle production generically to dominate compared to GPP if $q_*\gg k_R$.

\medskip
Another model-independent contribution to the abundance arises from gravitational freeze-in from the SM thermal bath, which is active if $T_R>M$.
We find~\cite{Redi:2020ffc}
\be
\Omega_{\rm DM}|_{\rm GFI}\approx 4\times  10^{-5} \frac{M k_R^3}{3 M_{\rm Pl}^2H_0^2}~,
\ee
where $k_R\approx T_R T_0/M_{\rm Pl}$ is again the comoving momentum that re-enters the horizon at reheating. If $q_*\lesssim k_R$, i.e. the modes re-enter during radiation domination,  gravitational freeze-in is  larger than eq.~\eqref{eq:Omega_sto} unless $\Delta(q_*)\approx 1$.
However, for $q_*\gtrsim k_R$, stochastic particle production is generally larger, and in the same regime it also dominates over GPP. 

\bigskip

To summarize there are regions of parameter where stochastic particle production is the dominant mechanism of production of DM especially if the 
the power spectrum is enhanced compared to the CMB value. Even for $\Delta(q_*)\approx 10^{-9}$, the DM abundance could be reproduced for large masses, $M\approx 10^{13}$ GeV.

\medskip
\paragraph{\bf Conclusions.--}
In this work, we introduced a new unavoidable mechanism of DM production from curvature perturbations in the early universe. 
These inhomogeneities break the conformal invariance of massless fermions, allowing particle production even in the massless limit. 
This should be contrasted with particle production in an expanding background, which is determined by the mass of the particle
and vanishes for $M=0$. At a technical level our results have been derived by studying the Bogoliubov transformation induced by the background, and by exploiting the stochastic nature of the source.

For perturbations produced during inflation, the DM abundance and distribution is fully determined by an integral of the power spectrum 
of curvature perturbations with only mild dependence on the cosmological history.
While this production mechanism is always present, it becomes quantitatively significant and relevant if the power spectrum is large for modes that are produced
towards the end of inflation. In particular, we find that for $\Delta(q_*)\approx 10^{-3}$ and $q_*\gtrsim 10^{-7}$ eV, 
this effect is larger than both gravitational freeze-in and particle production due to the mass. The DM abundance is reproduced for $M\gtrsim 10^6$ GeV.

Large inhomogeneities that lead to particle production are reminiscent of production of primordial black holes production from inflation,
leveraging the fact that the power spectrum is very poorly known at small scales. The black holes produced in 
the present scenario are very small and evaporate immediately not contributing to the DM abundance. 
They could however also contribute to the abundance of particles produced. 

We envision many generalizations of this work. In a companion paper, we will study particle production 
from stochastic backgrounds for fields of different spins, including scalar and tensor perturbations. The same effect will
also apply to dark sectors made from strongly coupled conformal field theories, possibly described by holographic duals. 
For interacting dark sectors, the mass could be generated dynamically from confinement in QCD-like dark sectors
modifying the results of particle particle production. We will return to a detailed phenomenological analysis of stochastically 
produced dark sectors in future work.

\medskip
\paragraph{ \small Acknowledgements.}
{\small  
AT thanks the Fermilab theory group for hospitality and partial support during the completion of this work.
}


\bibliographystyle{apsrev4-1}
\bibliography{letter_biblio}

\end{document}